\documentclass[12pt]{article}
\usepackage{epsfig}
\begin{document}

\centerline{\bf  High $Q^2$ HERA Events and pQCD at High x}
\centerline{  Stephen Rock and Peter Bosted}
\centerline {American University, Washington D.C. 20016}
\centerline{ June 1997}
\vskip .3in
\begin{abstract}
 We compare data on $F^p_2$ from SLAC experiments in the range 
$0.7\leq x \leq 0.97$ in and
near the resonance region with previous empirical fits
to Deep Inelastic Scattering and with calculations from parton
distribution functions.  The data is in rough agreement with the
empirical fits, but is an order of magnitude higher than the pQCD
calculations at the highest $x.$ This compares with the two orders of 
magnitude increase in the quark distributions at high $x$ which seem 
to be necessary to explain the HERA high $Q^2$ events. 
\end{abstract}
\vskip .3in

 Recent interest in the anomalous high $Q^2,$ high $x$ events at HERA 
\cite{HERA} has generated interest in the accuracy of the pQCD evolution 
to these
regions.  This evolution depends on $\int^1_x$ of the quark distributions,
thus making the high $x$ quark distributions important.
pQCD fits have ignored  data in the region $x\geq 0.75$ because of fear of 
possible higher twist contamination. However, there is a wealth of SLAC data
in the region up to $x=0.97$ \cite{SLAC,Whitlow},
albeit in the resonance region $(W^2\leq {\rm 4 GeV^2})$ and at relatively
low  $Q^2.$  Bloom-Gilman \cite{BG} duality states that the average of the
data in the resonance region approximates the DIS structure.  With that in
mind, we compare pQCD and various fits to the DIS region with the SLAC 
high-$x$ resonance data.

 Figure 1 shows a sample of the SLAC measurements of $F^p_2$  for $x\geq0.7$ and
 $7\leq Q^2\leq 30.$ $F^p_2$ decreases by approximately 3 orders of magnitude
over this $x$ range
and there is a $Q^2$ dependence of approximately a factor of 2 to 4 between
the various bands of data.

  Figure 2 shows the ratio the same SLAC data to 
the NMC DIS fit \cite{NMC}.
This fit used only data with $x\leq0 .75$. The data in Figure 2. 
span the range 
$7\leq Q^2\leq 30$.  In the range of validity of the fit ( $x\leq0 .75$), 
the ratio is
similar to unity, as expected. The ratio raises to about 2, at the highest $x$.
This fit is dominated by
a $(1-x)^3$ behavior at high $x$ as predicted by the quark counting 
rules \cite{Brodsky}.
The agreement is remarkable considering that the data at the highest $x$ is
in the region of the first resonance, far from the normal DIS region.
 The SLAC global fits used data up to x=0.85.
 As seen in Figure 3, similar good agreement is obtained when using the  
SLAC-$\Lambda_{12}$ \cite{Whitlow} as the denominator. This fit is also 
dominated by
a $(1-x)^3$ behavior at high $x.$
As observed in Figure 4, the ratio of data to 
the  SLAC-$\Omega_9$ fit  is close to unity for the high $Q^2$ data, but
the ratio is less than unity for the lower $Q^2$, high-$x$ points.
 Thus previous fits to world DIS data can  be used to 
estimate the $F^p_2$ structure function in the resonance region to within
a factor of about two.

 Figure 5 shows the ratio of SLAC data to a typical pQCD evolution fit,
CTEQ 4M \cite{pdf}. Again, there is good agreement for $x\approx0.7$.  However,
the ratio raises sharply and is about a factor of 10 at the highest $x.$
Note that since the data shown in Figure 1 decreases by a factor of 1000 over this
range in $x,$ this indicates that the pQCD fits decrease by 4 orders of
magnitude. 
 The difference between the pQCD fit and the data   
could be explained by ``higher twist'' effects, but they would
have to be an order of magnitude greater than the pQCD evolution and
also behave similar to the quark counting rule prediction of $(1-x)^3$.
 A typical higher twist form is 
$F_2^{pQCD}[1 +C_{HT}/(Q^2(1-x))]$ \cite{htBrod}.
Figure 6 shows the ratio of the data to this higher twist form with
$C_{HT}=4.$ While the rapid rise at high $x$ from pQCD alone is 
considerably suppressed, the DIS region of excellent pQCD fits 
($x\approx 0.7$) is suppressed 
by an unacceptably large 30\%.  A previous higher twist 
analysis \cite{htBCDMS} 
for $x\leq 0.75$ determined that the equivalent of $C_{HT}$ is $\approx 0.3,$
an order of magnitude smaller than that used in Figure 6..
Of course, more complex higher twist terms could
be included with higher powers of $Q^2(1-x)$ which would eventually fit the
data.

 Recently Kuhlmann, Lai and Tung \cite{Tung} have tried to explain the HERA events using
a pQCD toy model which enhances the $u$ quark probability beyond the CTEQ4M
fit by approximately two orders of magnitude at x=0.95.  This pQCD
fit would overestimate the SLAC data by an order of magnitude and thus
require a huge negative ``higher twist effect'' of approximately the form
discussed above.
 
  In conclusion, the SLAC data in the kinematic region $.75\leq x \leq .97$
which is mostly in the resonance region, is an order of magnitude higher
than current pQCD parton distributions , and an order of magnitude less than 
the toy models created created to explain the high $Q^2$ HERA events. Existing
DIS fits with quark counting rule behavior at high $x$ are a reasonable
approximation of the data, consistant with Bloom-Gilman duality.
A dedicated experiment with the SLAC 50 GeV electron beam could measure
$F^p_2$ in
the kinematic region $0.80\leq x \leq 0.95$ in the region of 
$Q^2\geq 50{\rm GeV^2}$ with the possibility of tagging charm.
This data could be used to disentangle higher twist effects from pQCD
evolution, and determine the $u$ and $c$ quark distributions at high $x.$

\pagestyle{empty}
\begin{figure}
\vspace*{7.7in}
\hspace*{.45in}
\includegraphics{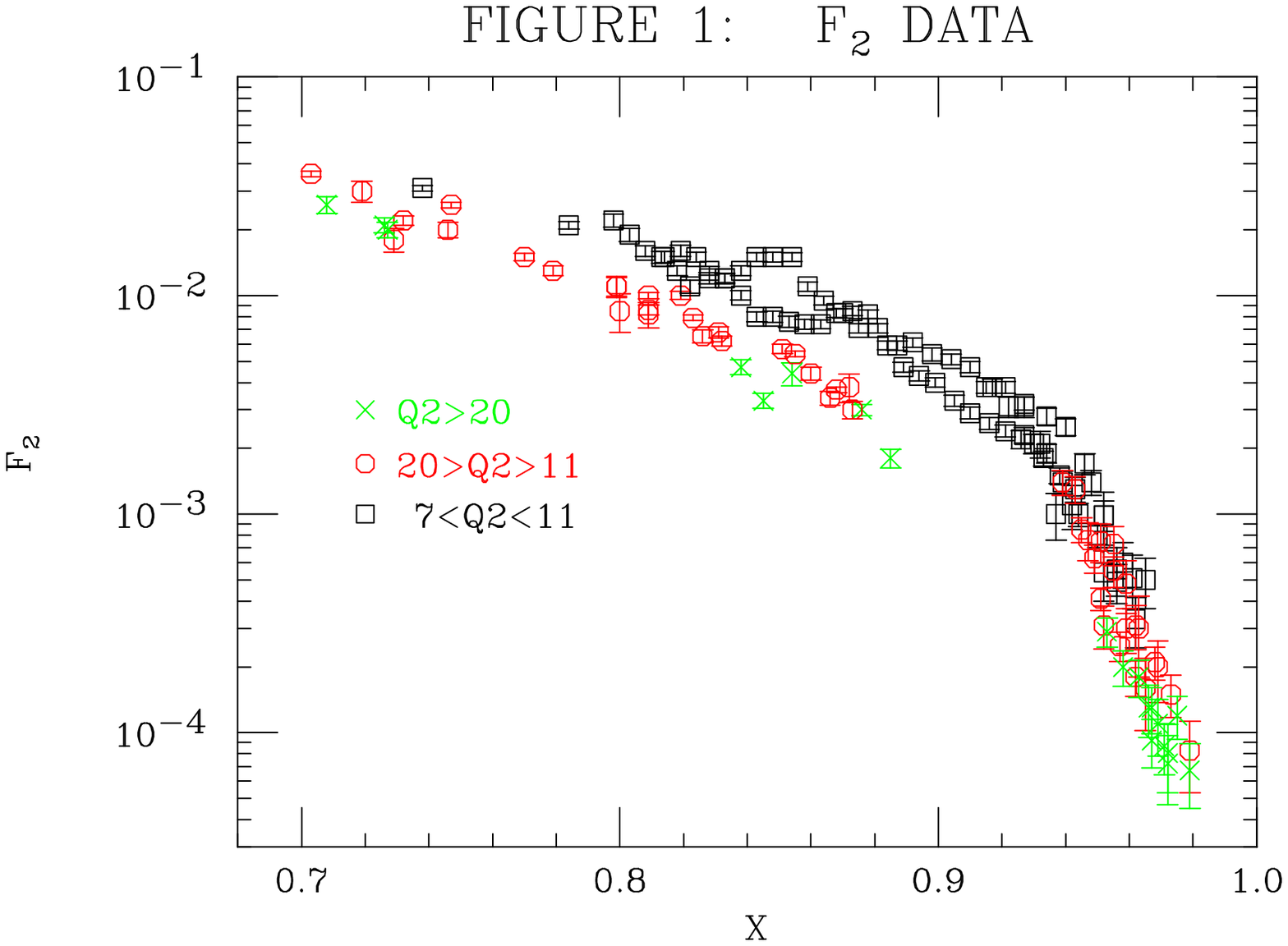}
\caption{ Selected SLAC data for $F^p_2$ in the range $0.7\leq x \leq 0.97$  and
$7\leq Q^2 \leq 30.$ Most of the data at the higher $x$ values is in the
resonance region.}
\label{a}
\end{figure}

\begin{figure}
\vspace*{7.7in}
\hspace*{.45in}
\includegraphics{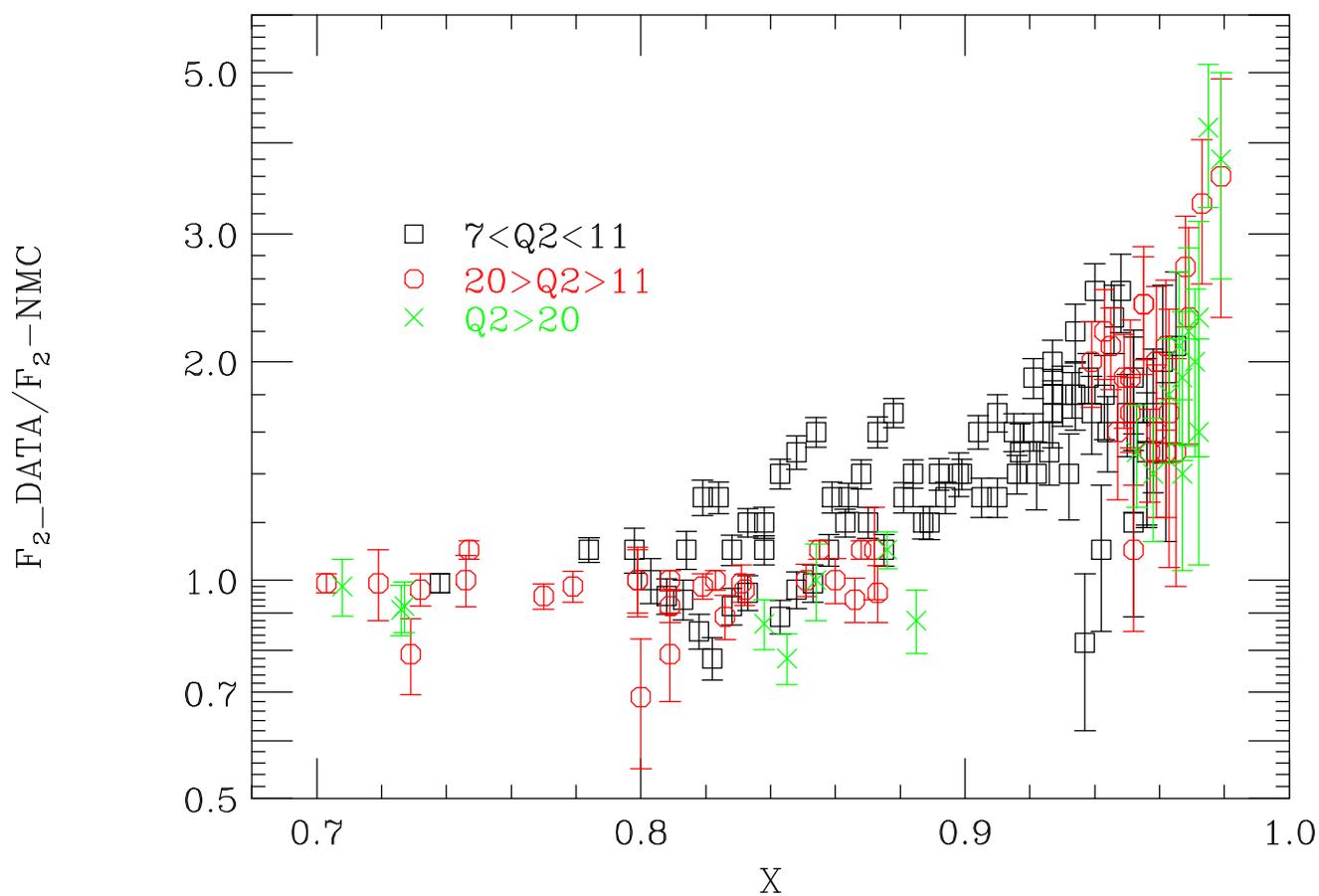}
\caption{ The data of Figure 1 divided by the NMC fit to Deep Inelastic 
Scattering Data for $x\leq 0.75.$}
\label{b}
\end{figure}

\begin{figure}
\vspace*{7.7in}
\hspace*{.45in}
\includegraphics{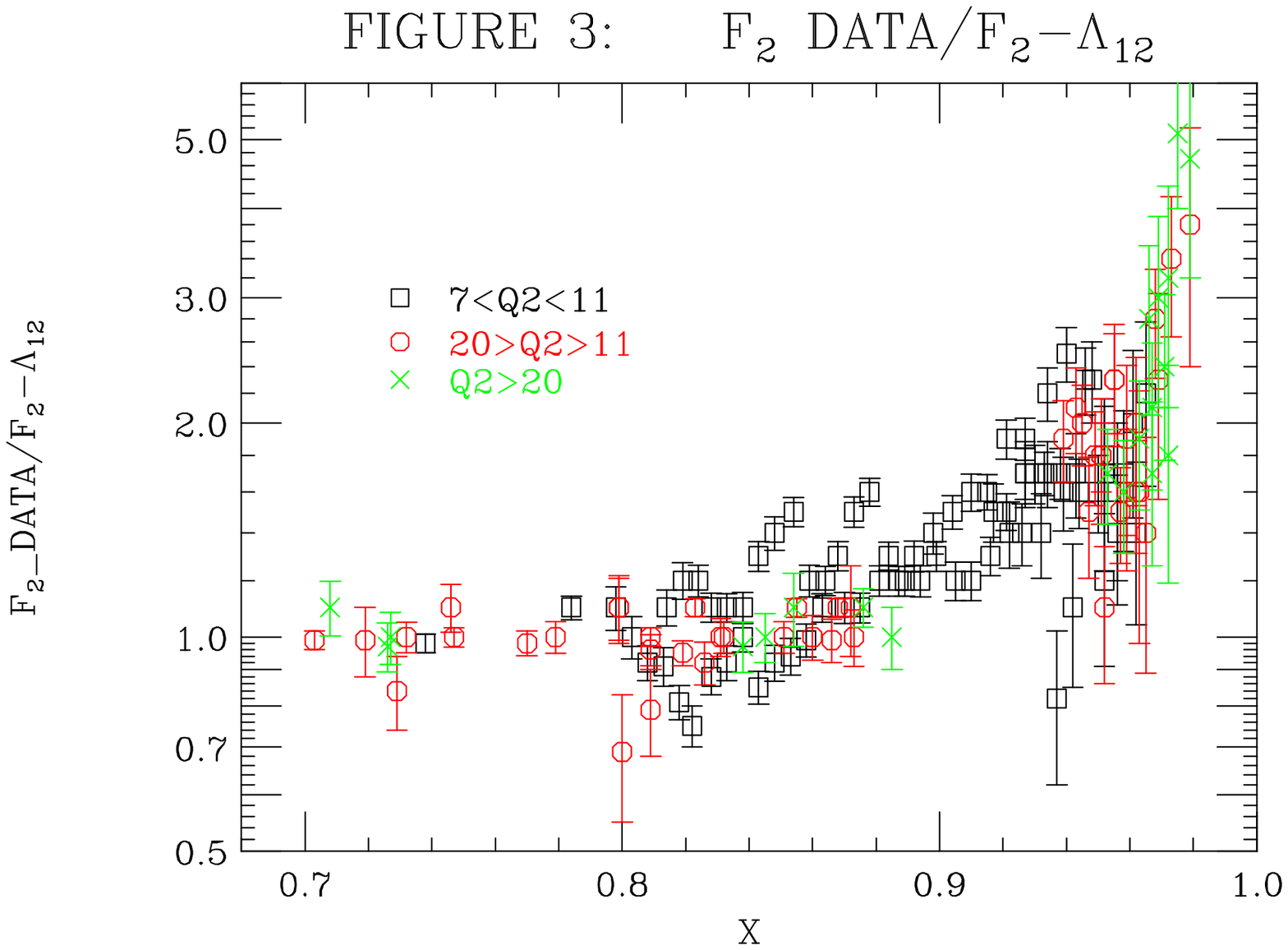}
\caption{ The data of Figure 1 for $F^p_2$ divided by the SLAC Global Fit 
$\Lambda_{12}$ to  SLAC Deep Inelastic Scattering Data for $x\leq 0.85.$}
\label{c}
\end{figure}

\begin{figure}
\vspace*{7.7in}
\hspace*{.45in}
\includegraphics{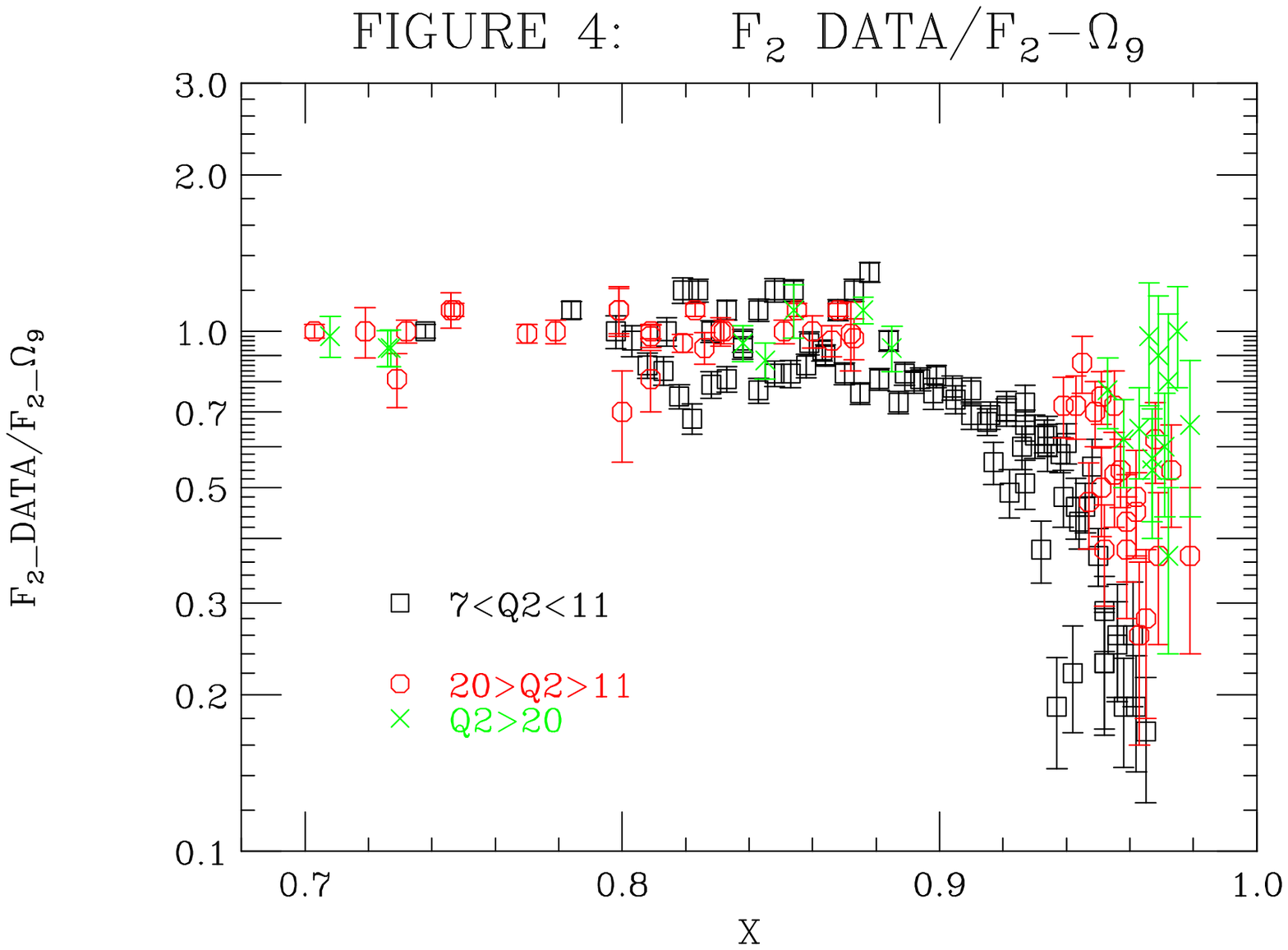}
\caption{The data of Figure 1  for $F^p_2$ divided by the SLAC Global Fit 
$\Omega_{9}$ to  SLAC Deep Inelastic Scattering Data for $x\leq 0.85.$}
\label{d}
\end{figure}

\begin{figure}
\vspace*{7.7in}
\hspace*{.45in}
\includegraphics{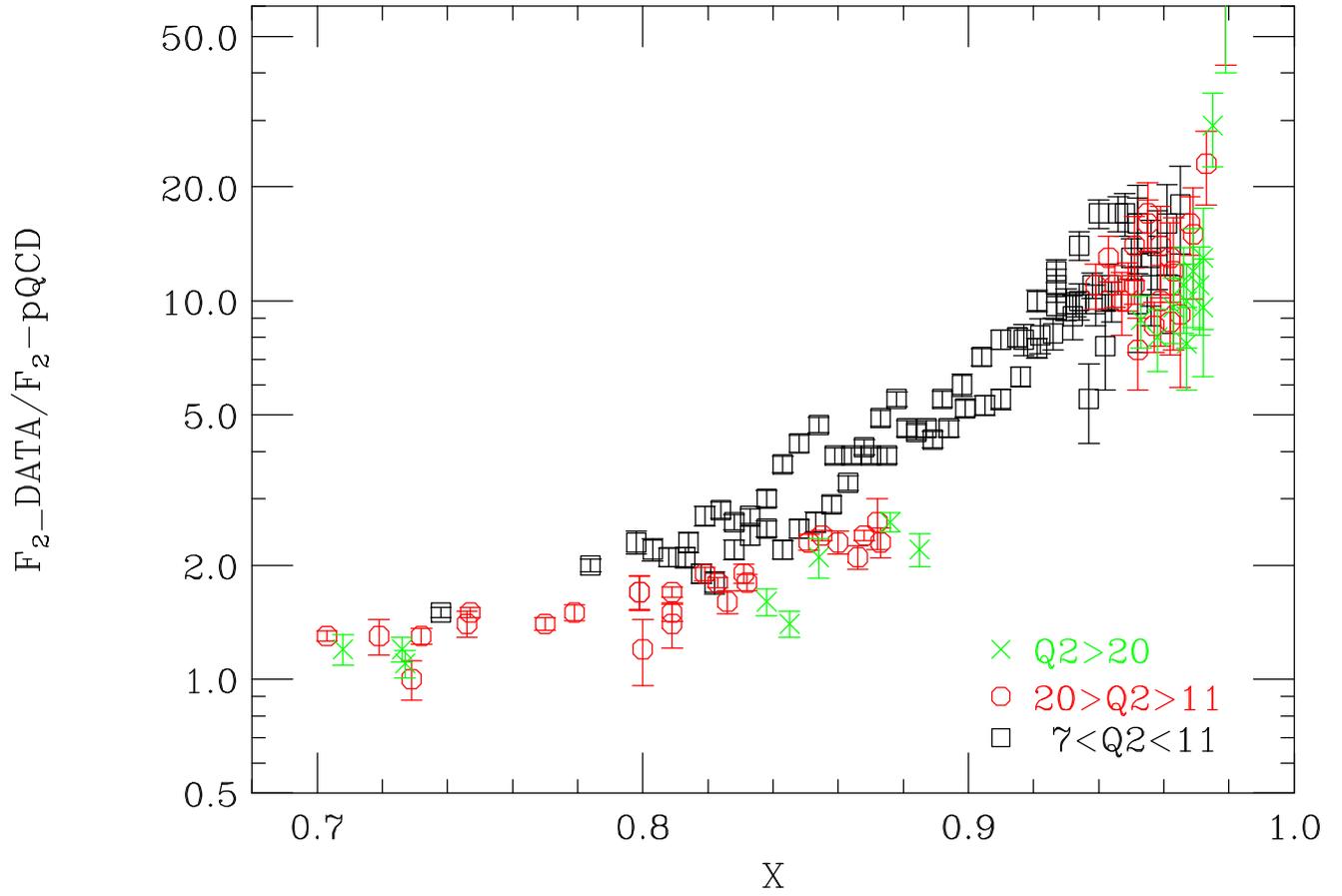}
\caption {The data of Figure 1  for $F^p_2$ divided by the pQCD NLO fit 4M 
done by  CTEQ.}
\label{e}
\end{figure}

\begin{figure}
\vspace*{7.7in}
\hspace*{.45in}
\includegraphics{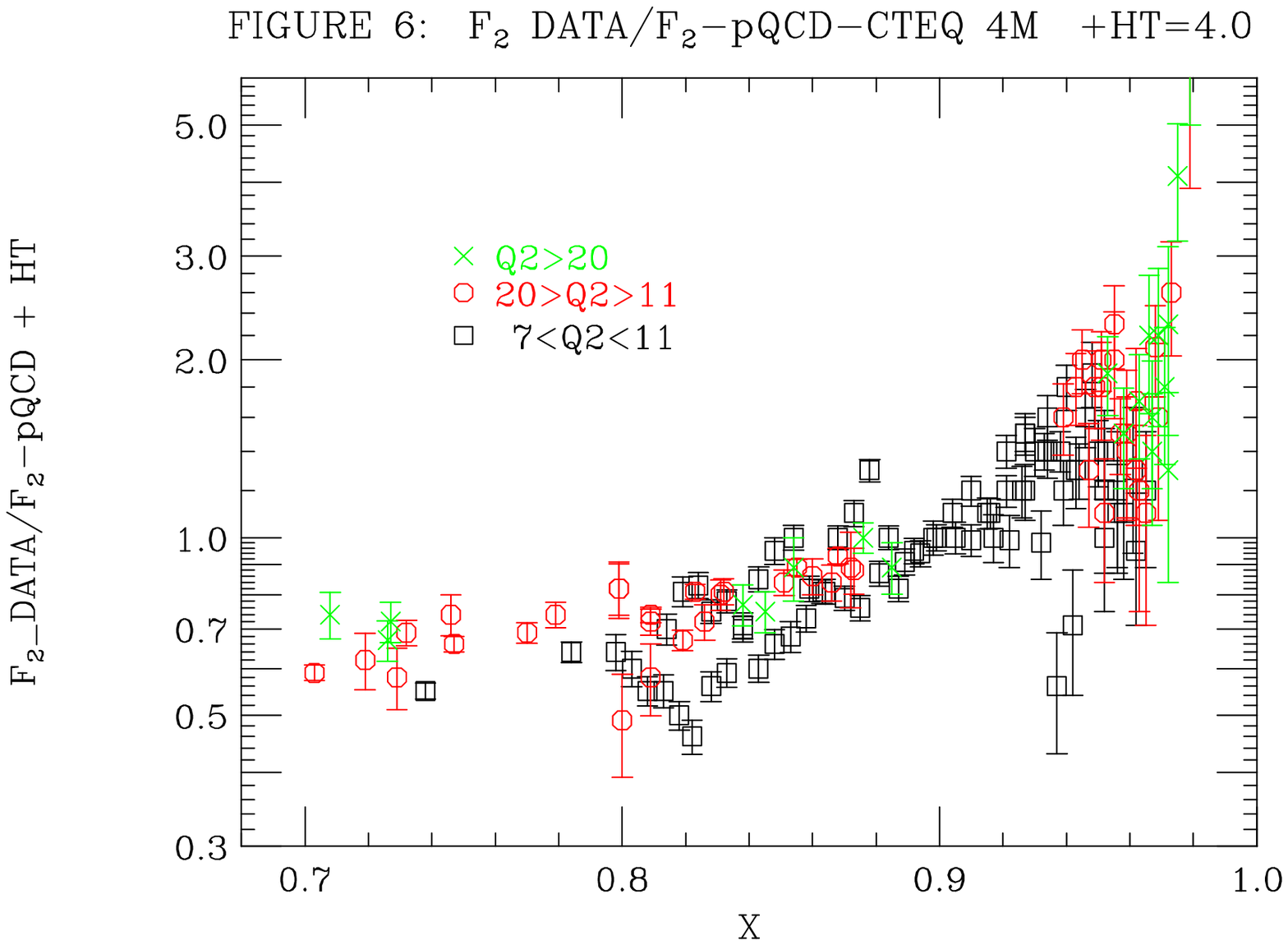}
\caption{ The data of Figure 1  for $F^p_2$ divided by the pQCD NLO fit 4M done by  CTEQ with
an additional empirical higher twist term described in the text.}
\label{f}
\end{figure}

\end{document}